# Investigating students' scientific reasoning through heuristic and analytical thought processes

Dimitrios Gousopoulos

Department of Psychology, National and Kapodistrian University of Athens, School of Philosophy

In recent years there has been growing evidence that even after teaching designed to address the learning difficulties dictated by literature, many physics learners fail to create the proper reasoning chains that connect the fundamental principles and lead to reasoned predictions. Even though students have the required knowledge and skills, they are often based on a variety of intuitive reasoning that leads them to wrong conclusions. This research studies students' reasoning on science problems through heuristic - analytical thought processes (System 1 - System 2). System 1 operates automatically and quickly with little or no effort and no sense of voluntary control, while System 2 focuses on the demanding mental activities that require it and is slow based on rules. Specifically, we seek to study those cognitive processes and information available to students when they face science problems and, therefore, to explore the various heuristic processes that students use when solving physics problems. Our results indicated four intuitive heuristics in students' minds when they solve problems in Mechanics and especially in the unit projectile motion: associative activation, processing fluency, attribute substitution and anchoring effect. These heuristics prevent students from applying knowledge and methods that they already possess to solve a physics problem.

## Introduction

Students in physics classes often fail to create the proper reasoning chains that connect the fundamental principles, and they generate shallow responses to the questions and problems they face (Meltzer & Thornton, 2012). The science education community has developed targeted instruction to address student conceptual and reasoning difficulties. However, in many contexts, these interventions did not lead to significant improvements in student performance, even though they possess the required knowledge and skills (Close & Heron, 2010; Kryjevskaia et al., 2011). More specifically, an emerging body of evidence suggests that students abandon the analytical way of approaching problems and rely on a more intuitive way of thinking that often leads them to wrong answers. (Kautz et al., 2005, Kryjevskaia et al., 2014).

In recent years, science education researchers have tried to shed light on the intuitive way students think using the dual-process theories of reasoning and decision making. For example, education scientists have explored and analysed the heuristics used by students to rank chemical compounds based on acid strength (McClary & Talanquer,

* dimgouso@primedu.uoa.gr

2011), while others have examined students' intuitive thinking in physics topics such as force (Wood et al., 2016) and buoyancy (Gette et al., 2018). The dual-process theories have provided insight into how people reason under conditions of limited time, knowledge and computational power (Shah & Oppenheimer, 2008; Kahnemann, 2011), which are similar to those faced by our students in a physics classroom.

In this study, we try to shed light on observed patterns of students' reasoning through the lens of heuristic and analytical thought processes. The topic of projectile motion is used as a context of investigation. More specifically, we examine the heuristics that secondary students use when trying to answer questions that demand making comparisons between the properties of two systems and problems that demand visualizing the situations and events by sketching an image, translating word to algebraic expressions, identifying the physics concepts and the target quantities that are useful to reach a solution, selecting the formula for the target quantities and solving multi-step equations.

## Theoretical Framework

It is a common expectation that, after the appropriate educational intervention, students will consciously and systematically construct chains of reasoning that start from established scientific principles and lead to reasoned predictions. When students' performance on course exams does not reveal such models, it is often assumed that students either lack the proper understanding of relevant scientific knowledge or cannot construct such chains of reasoning (Heckler, 2011). Psychological research in the areas of thought and reasoning, on the other hand, suggests that, in most cases, thought follows paths quite differently from those previously mentioned. The theory of the two systems of the mind suggests that there are two different processes involved in many cognitive tasks, known as System 1 (Heuristic) and System 2 (which contains the analytical procedures). The first system supports fast, intuitive, and automatic thoughts, while the second system is slow and rule-based. Some examples of automatic actions that are attributed to System 1 are: "Determining that one object is further away than another", "Answer to the question 2+2=?", whereas examples of actions that are attributed to System 2 are: "Search for a woman with white hair", "Comparison of the selling price of two devices".
In many cases, these two mind systems produce different answers. According to Evans (2006), when a person confronts a particular situation or a problem that has to be solved, System 1 directly and subconsciously develops a mental model of the situation based on past knowledge and experiences, relevant facts, and other factors. This "first available" mental model often represents the person's quick and subconscious effort to produce a coherent and reliable way to approach the given situation or problem. According to Evans, the mind creates one model at a time (singularity principle). The model that results from the heuristic processes of the mind is based on the person's most recent knowledge and beliefs. Once the first available mental model is developed, it becomes available for testing from the most rigorous and analytical System 2. However, if the individual feels confident with the first available mental model, the analytical process may be completely overridden. If this happens, then the first available model produces the final answer. This immediate transition from the first sense to the final answer (without necessarily the intervention of System 2) prevails in our daily activities. A characteristic example is the following: Due to the coincidence of 2 planes going down last month, many people may prefer to travel by train. In fact, the risk has not changed. In general, System 1 is quite efficient and accurate in providing quick assessments of known situations. Also, even if System 1 proposes a wrong conclusion or a wrong action in a field of everyday life, often we are not able to realize the failure of these conclusions since we receive no direct feedback for our mistakes. Because of this, it is not surprising that we learn to trust our intuition, which seems to us to be efficient in our everyday life. This leads novice physics students to transfer their confidence to intuitively appealing thoughts from the everyday context within science, and it does not often need the explicit and rigorous validation of their thinking. To identify an error of intuitive thinking, System 2 should be involved. If analytical thinking is not satisfied with the current System 1 response, then the reasoning path returns to System 1 to produce a new mental model to be examined (by System 2). It is important to stress that System 2 has its own bias. Therefore, the involvement of analytical procedures does not necessarily ensure the detection of a defect in the sequence of thinking, or the production of an answer based on logical arguments. Contrary to sound scientific thinking, individuals often fail to search spontaneously for alternative mental models or for items that might falsify their original predictions. Therefore, even when the analytical process is engaged, the first impression, the intuitive answer often appears as the final answer (Frankish, 2010; Thompson et al., 2018).

A rational judgment requires replacing the intuitive and automated way of thinking with an effective analytical way of thinking. System 2 interventions may inhibit or modify the responses generated by System 1 processes depending on the degree to which the response is considered unsatisfactory. These interventions are more likely to occur when people have developed metacognitive skills and have a high cognitive ability. System 1 processing is expected to prevail when an individual has less knowledge, skills, or motivation to work (Talanquer, 2014).

Many of the intuitive processes in System 1 can be thought of as shortcut strategies, which are called heuristics and reduce the information and data processed by the mind. In general, heuristics simplify a person's reasoning by reducing the number of elements used to make a decision or by providing implicit rules for how and where to look for information, when to stop searching, and what to do with the results (Gilovich et al., 2002; Kahneman, 2011; Shah & Oppenheimer, 2002). For instance: If an executive director wants to answer the question, “Should I invest in Ford stocks?”, he might replace the previous question with this, “Do I like Ford’s cars?”.
In the following sections, we will discuss, in turn, four heuristics which were frequently used by our study participants.

**Processing Fluency**

A problem or a situation includes pieces of information that are not all processed with the same ease by the mind. For naive learners, clear features are easier to access than the most tacit ones. Processing Fluency is related to the processing time of the features involved in the problem. According to it: the faster the mind processes an idea, the more "weight" is given to it when creating a reasoning process (Morewedge & Kahneman, 2010; Oppenheimer , 2008).

Information that is easier for a student to observe and process is expected to influence his or her response to a given problem, especially if such information can be correlated in some way with the targeted variables of the problem.

**Attribute Substitution**

People can quickly generate intuitive answers to difficult questions and complex issues. A suggested explanatory mechanism is the following: If a satisfactory answer to a challenging question is not quickly found, System 1 will discover and answer an easier relevant question. The function of answering one question in place of another is called substitution. According to Kahneman (2011), the question-goal is the assessment the individual intends to produce, while the heuristic question is the simplest question to be answered instead of the question-goal. For example, when students are asked what the distance is between the point of fire of a missile and the point of strike with the ground, some students may substitute the above question - target with the following heuristic question: What is the range of the projectile?

**Associative activation**

Associative processing uses connected structures of the mind to supplement information quickly and automatically in situations resembling previous experiences or observations. In general, judgment and decision-making involve weighing various pieces of information. Cognitive biases arise when some aspects of information are systematically given more weight while others are degraded or neglected. Research suggests that highly activated information is likely to be more important than it is worth. In contrast, relevant information that is weakly activated or not activated at all will eventually be neglected by the mind. In general, decision-making is based on existing and "on-the-spot" correlations between the information that has been activated. (Morewedge & Kahneman, 2010).

The aforementioned heuristic processes are based on associative processes, and often these heuristics appear in combination to construct the first available intuitive model for interpreting a given situation.

**Anchoring Effect**

System 2 works with data retrieved from memory through the automatic and unintentional operation of System 1. Thus, System 2 is prone to the influence of anchors that make it easier the retrieval of certain information. System 2 has no control over this phenomenon and is unaware of it. Individuals give much more "weight" to a piece of initial information. This piece of information is called an “anchor”.

Research in science education shows that preconscious and domain-specific cognitive elements are essential in guiding and constraining students’ scientific reasoning. Such cognitive elements are p-prims (diSessa, 1993), implicit presuppositions (Vosniadou, 1994) and implicit assumptions (Talanquer, 2006). In contrast with these elements, few researchers have examined a more domain-general implicit students’ reasoning strategies (Stavy & Tirosh, 2000; McClary & Talanquer, 2011). For example, the intuitive rules theory of Stavy and Tirosh includes heuristics of More A – More B that can describe students’ biases in comparison tasks. Furthermore, a few studies seek to identify heuristics that students use when facing a problem in science. More specifically, McClary and Talanquer (2011) have identified the heuristics when students rank chemical compounds based on acid strength. In the area of Physics and particular in the topics of capacitors and waves Kryjevskaia et al. (2014) have identified the availability heuristic in students’ reasoning. Gette et al. (2018) and Heckler et al. (2018) have reported identical findings when seeking to identify students’ heuristics in the topics of buoyancy and mechanics (projectile motion, sliding, balance scale etc.).

**The originality of the research**

Recent research has shown that errors in student responses are not necessarily due to a lack of conceptual comprehension but to the inability to generate appropriate links between the suitable scientific structures so that the constructed mental model used to approach the problem is scientifically correct. Educational research that aims to identify key mechanisms leading to observed patterns of thought is minimal, even though cognitive scientists have developed different models of interpretation of human thought and decision-making.

Although the theories of the dual system of the mind are widely used in many areas of psychology (decision-making, motivation, etc.), they have been mainly ignored by scientists conducting educational research. In a relatively recent survey, Heckler (2014) underlined the applicability of theories related to the two mind systems to a deeper analysis of students' thoughts when seeking to solve scientific problems.

Research on heuristic thinking processes has taken place in non-academic / school contexts. However, there are indications that these processes (System 1) play a central role in the classroom, which is of great value to be studied. Heuristic thinking processes such as heuristics of availability have been formulated by scientists honoured with the Nobel Prize for their overall work in mental processes and decision-making. This project has been implemented in many areas of both psychology and economics. Based on this award-winning work, we seek to apply the observed experimental findings and conclusions to the students' reasoning when confronting scientific problems in the natural sciences. Any relevant effort to implement the previous findings has been minimal and is only within the framework of the university class.

Based on those mentioned above, we consider the proposed research extremely useful, contributing to the overall effort of the research community to interpret and map out mental processes during the solution of scientific problems and situations.

## Research Design

The central aim of our research proposal is to detect the intuitive heuristics students use when solving problems related to projectile motion.

The research question is: What heuristics do 11th Grade students use when solving problems in Mechanics (projectile motion)

### Participants

Forty-five students (20 female, 25 male) of 11th Grade were selected based on their performance in a preliminary inquiry that tested their basic knowledge on the thematic unit of "Projectile Motion". In order to disentangle students' knowledge from reasoning strategies, we constructed a test with screening questions in the preliminary phase of our study since we aimed to probe the students' tendency to abandon the appropriate formal reasoning in favour of a more intuitive and appealing approach (Gette & Kryjevskaia, 2019). Approximately 90% of the initial students answered the screening question correctly and constituted our sample. These students were taught for five weeks the thematic unit "Projectile Motion" according to the official instructions of the curriculum of the Greek Ministry of Education. In addition to a detailed presentation of the theoretical concepts, the instructional process included many applications, exercises, and problems.

### Research Instrument

We designed a research instrument based on questions and problems that students commonly encounter in high school physics textbooks (Appendix) and on prior research on heuristic reasoning. In particular, the first three questions require students to compare the properties of two objects thrown horizontally. For some students, question A1 elicits an intuitive idea that "If I double the height, the distance travelled doubles as well", question A2 elicits the well-documented idea that "More velocity – Less time", and question A3 elicits the idea that "More mass – Less time". Questions B1 and B2 require students to recall and apply the basic formulas in projectile motion. Question B3 tends to elicit a strong intuitive response that velocity has only one component, the speed value will its direction is usually ignored. The last question, B4, is an interesting question that requires students to be careful regarding the change in a requested time interval. This question elicits the tendency to rely too heavily on the first piece of information offered (the "anchor") when making decisions or calculations. In this specific question, the time interval plays the role of the anchor.

The first three problem's questions are related to the time interval that the body is required to reach the ground (t =2s), whereas the last problem's question relates to a different time interval (t=1s). Before the survey, brief interviews were conducted with students, and all questions were tested so that participants could understand the given information. Moreover, three experienced Physics teachers (over 25 years of experience) checked the given questions for their scientific accuracy and participated in the initial formulation and selection of the questions.

### Data Collection

Students completed the research instrument during a 30-minute semi-structured interview. During the interview, participants were asked to think aloud as they solved each question. For each question (A1 to A3 and B1 to B4) they were given a time limit of 2 minutes (Maeyer & Talanquer, 2010; Gillard et al., 2009b; Kelemen & Rosset, 2009) so as the more intuitive responses to emerge. Moreover, they were not allowed to review the previous questions. Their responses were collected in written form using a worksheet, and their verbal explanations were audio-recorded for further analysis.

### Data Analysis

Interview transcripts were reviewed and summarized for each participant and question. All the individual answer summaries were then analyzed using an iterative, constant comparison method of analysis in which common reasoning strategies were identified within each answer category (Maeyee & Talanquer, 2010; Charmaz, 2006). We identified different patterns of reasoning for each answer group and then compared them to the patterns of reasoning followed by students who yielded different answers for a particular question. The final coding system was developed in such a way so as the two coders reached an approximately 90% agreement. During the data analysis, codes and task responses' interpretation were continually revised for the reasoning paths to better identify within each task.

## Findings and Analysis

Our analysis led us to identify four main heuristics utilized by our study participants when trying to answer questions related to the projectile motion of an object. Some heuristics are similar to those reported in previous research (Kryjevskaia, Stetzer & Grosz, 2014; Gette et al., 2018). The main heuristics found are associative activation, processing fluency, attribute substitution and anchoring effect.

*Associative activation*

Associative processing uses linked structures to answer problems and situations with similar features to past observations or experiences. Many of these associations take the form of direct (More A – More B) or inverse (More A – less B) correlations built upon activated knowledge (Talanquer,2014). In question A1 activating the idea of height in the context of projectile motion seems to activate an incorrect analogical relationship between height and range in students' minds. After analyzing the students' answers, we noticed that 29/45 (64.4%) of them followed this reasoning. For example:

"Since we launched the body from twice the height, its range will double its value."

Even when System 2 intervenes (through its analytical approach), the coherence of height and range in a linear-proportional way is so strong for some students that they eventually choose to ignore the findings of System 1. An interesting finding that supports this statement is that 9/29 (31%) students approached the question correctly, and one step before the solution of the problem they abandoned the analytical way of thinking and followed a more intuitive way and finally answered incorrectly activating the proportional correlation of the two variables. Their justification was that they were sure throughout the problem solution of the proportional relationship between the range and the height. When they got to a point where they had to divide the square root of 4 with the square root of 2, even though they knew how to do this, they chose to divide wrong to confirm their initial reasoning. A typical example of this was the approach of Participant S4:

$$\left.\begin{aligned} s_2 &= v_0\sqrt{\frac{4h}{g}} \\ s_1 &= v_0\sqrt{\frac{2h}{g}} \end{aligned}\right\} \Rightarrow \frac{s_2}{s_1} = \frac{\sqrt{4}}{\sqrt{2}} = \sqrt{2} \Rightarrow s_2 = 2s_1$$

S4: “I must have made a mathematical mistake somewhere, but I am pretty sure that the range doubles its value.”

*Processing Fluency*

Processing fluency is a cognitive bias in which the cues that are easier to notice and process by students are expected to influence their responses in each task they face. According to processing fluency, when students are in a state of processing fluency, they like what they see, trust their intuitions and feel that the current state is familiar. After the data analysis regarding question A2, we noticed that 12/45 (26.6%) students have greatly emphasised the initial body velocity. The students' attention to this variable probably led to an erroneous correlation between the initial velocity of the body and the time the body needed to reach the ground (i.e., associative activation). Therefore, many students answered that the body with twice the initial speed would arrive either sooner than the other body or later. The heuristic of processing fluency describes the above characteristics in conjunction with the heuristic of associative activation.

Interviewer: Can you find any other quantities that could possibly affect the outcome?
S21: Um, no!

“The object with the greater speed will arrive quicker.”

Furthermore, after analysing the students' answers regarding question A3, we noticed that 8/45 (17.7%) students seem to have greatly emphasised body mass. These students paid great attention to this variable, which probably led them to an erroneous correlation between the body mass and the time of arrival needed for the body to reach the ground (i.e., associative activation). Therefore, the students answered that the body with twice the mass would arrive either sooner than the other body or later. The heuristic of processing fluency describes the above characteristics in conjunction with the heuristic of associative activation.

.
“The heavier object will feel a greater drag and arrive later.”

*Attribute Substitution*

According to this heuristic, if a satisfactory answer to a difficult question is not found quickly, System 1 will find an easier relevant question to answer. Question-goal is the assessment that the individual intends to produce, while heuristic question is the simplest question that is answered instead. In the third question of the problem given, which concerns the calculation of the body velocity when it reaches the ground, many students (10/45 - 22.2%) replaced the question with the question: “What is the body speed?” without including the direction of the velocity in their answer. This can be interpreted as follows: Velocity, as a vector quantity, has two characteristics, its speed value and its direction. The measure of speed seems to be more activated in the mind of these students. As a result, the concept of velocity is replaced by the concept of speed since it is more available in their minds. The heuristic of attribute substitution can describe the reasoning mentioned above.

S21: So, the speed of the object is $10\sqrt{5}\, m/s$
Interviewer: Could you please repeat the question given to you?
S21 (without looking at the question): Um, yes, what is the body speed?

*Anchoring Effect*

Anchoring is a cognitive bias in which individuals give much more “weight” to an initial piece of information offered. This piece of information is called "anchor". An interesting finding is this: In the first three questions, students were asked to calculate variables that describe the body at the moment of its arrival on the ground (t = 2sec in this problem). In the last (4th) question of the problem given, students had to calculate a variable (the rate of change in the kinetic energy of the body) at the time t

= 1sec. A large percentage of students, while using the correct mathematical formula to find the rate of change of kinetic energy (36/45), replaced the time t in the mathematical formula with t = 2sec and not with t = 1sec. The previous event indicates that the students were "hooked" on the value of the time variable equal to 2 sec (16/36 - 44.4%). This anchor prevented them from activating the information that the time for which the rate of change of the body's kinetic energy was requested is 1 sec. This kind of reasoning can be attributed to the anchoring effect. Consider, for instance, the reasoning of Participant S13 while solving the last question of the problem:

S13: The rate of change in the kinetic energy of the body is equal to $mg^2t$. I substitute m for 1kg, g for 10 $m/s^2$ and t for 2 seconds.
Interviewer: What about the time?
S13: I calculate the rate of the time it (the object) reaches the ground (that is for t =2s)

Finally, after processing our research data, we noticed that the performance of the students in questions B1 and B2, which check the understanding of basic concepts and the application of the basic mathematical relationships of the horizontal shot, does not differ from each other, since of the 45 participants 44 answered both questions correctly. Interestingly, 65% of the students who showed evidence of cognitive ease in question A2 (20 students) also showed the same evidence in question A3 (13 students).

## Discussion

The trigger for this research was the observation that students who have the appropriate knowledge and skills in a subject are unable, in many cases, to apply them consistently to build scientifically correct inferences and solutions. These inconsistencies are persistent to questions that elicit strong intuitive responses. The current study aimed to identify cognitive mechanisms contributing to these inconsistencies. According to dual process theories of reasoning, students rely on several heuristics to make decisions and answer a given question. Our study revealed that the secondary students who participated used four major heuristics: *associative activation, processing fluency, attribute substitution, anchoring effect* to solve two types of questions concerning projectile motion. In particular, the research instrument included questions that require students to make comparisons between the properties of two systems, and questions that demand visualizing the situations and events by sketching an image, translating word to algebraic expressions, identifying the physics concepts and the target quantities that is useful to reach a solution, selecting the formula for the target quantities and solving multi-step equations. These heuristic strategies are similar to those reported in relevant previous work (Kryjevskaia, Stetzer & Grosz, 2014; Gette et al., 2018).
Even though participants had the appropriate knowledge and skills, they failed to apply them to questions that elicited intuitive responses, abandoning the formal analytical way of thinking. They relied on more intuitive reasoning strategies confirming the literature's relevant findings (Kautz et al., 2005; Kryjevskaia, Stetzer & Grosz, 2014).

Research has shown that the combination of attribute substitution with associative activation and processing fluency can give us the appropriate tools to interpret a significant portion of incorrect or naïve responses of college science students (Talanquer, 2014). Our study showed that heuristics act in conjunction when students face a comparison task or a problem requiring an analytical solution, confirming the findings of the previously mentioned research studies. More specifically, initial answers generated by some of the secondary students who participated in our study were triggered by the tasks' most explicit features, such as the initial velocity in question A2 and the object's mass in question A3. These features were the most accessible and easily processed (processing fluency) and led to the activation of related knowledge (associative activation). Moreover, the more strongly activated information, such as the analogical relationship between height and range in question A1, greatly influenced students's judgements and constrained their scientific reasoning. Besides, judgement biases can be described as overweighting some aspects of information and underweighting others (Kahnemann and Frederick, 2002).

Moreover, students automatically evaluate the associated characteristics when facing a target assessment. If one of these attributes is more accessible, it could be used as a substitute in making the decision that will lead to a specific solution to the query given (Morewage & Kahnemann, 2010). Respectively, some of the participants in our study replaced the object's velocity with the more accessible quantity of speed. This had, as a result, the participants to answer a simpler and more accessible question (i.e., "What is the speed of the body?") as a replacement for the response to the more difficult question (i.e., "What is the velocity of the body?"). Finally, people often rely too much on the first piece of information they are given about a topic. When making estimates about something, they interpret newer information from the reference point of their anchor. Respectively, a few of the secondary students who participated in the present study over-relied on the first information regarding the requested time interval of t = 2sec to prevent them from updating the new requested time interval and making a mistake during the solution to the question given.

Many of the heuristics described in this paper act in tandem or conjunction when students work towards the solution of a task given. This interconnection between them makes it difficult to propose targeted strategies to help students avoid these cognitive biases. According to dual–system models, the errors that people make during a cognitive task, such as problem-solving, are due to two failures: the automatic processes of System 1 generate a faulty intuition, which the controlled processes of the System 2 fail to detect. So, one possible way to help students is to better recognize the use of their heuristics in the tasks that they are engaged, when they apply them and how to exert control over their application.

Finally, additional research is required to explore the use of heuristics strategies in the context of different subjects and educational levels and how we can reform our curricula in science to enhance our students' scientific learning and teach them how to monitor and recognize the use of their heuristics, a skill that would lead them to make better decisions in their lives.

**Appendix**

The research Instrument.

**A1.** Two spheres 1 and 2 are launched simultaneously from points A and B which are in height h and 2h respectively, with the same horizontal initial velocity. The air resistance is considered negligible. If the range of the projectile motion of the spheres 1 and 2 are $s_1$ and $s_2$ respectively, then:

**a)** $s_1 = s_2$ **b)** $s_2 = 2s_1$ **c)** $s_2 = s_1 \cdot \sqrt{2}$

**A2.** Two bodies A and B are thrown simultaneously from the same height with horizontal velocities of $\upsilon_0$ and $2\upsilon_0$ respectively. Body A reaches the ground.

**a)** before body B

**b)** after body B

**c)** simultaneously with body B

**A3.** Two bodies A and B with masses $m_1$ and $m_2$ respectively are launched simultaneously from the same height h with equal horizontal velocities. The mass of body A is twice the mass of body B. If $t_1$ and $t_2$ are the moments when body A and body B reach the ground respectively, then

**a)** $t_1 = t_2$ **b)** $t_1 = 2t_2$ **c)** $t_2 = 2t_1$

**Problem B**

A small body of mass m = 2kg is thrown horizontally from a height h = 20m above the ground with an initial velocity $\upsilon_0$ = 10m / s. Calculate:

**1)** the time it takes for the body to reach the ground,

**2)** the horizontal distance traveled by the body until it reaches the ground,

**3)** the velocity of the body when it hits the ground

**4)** the rate of change in the Kinetic energy of the body at time t = 1s.

Consider g = 10m / $s^2$, and the air resistance negligible.